\documentclass[english,british,sort&compress]{article}
\usepackage[T1]{fontenc}
\usepackage[latin9]{inputenc}
\usepackage[a4paper]{geometry}
\usepackage{wrapfig}
\usepackage{amsmath}
\usepackage{graphicx}
\usepackage{esint}
\usepackage[numbers]{natbib}


\usepackage{babel}
\begin{document}
\selectlanguage{english}%
\global\long\def\mat#1#2#3{{#1}_{\hphantom{#2}#3}^{#2}}

\selectlanguage{british}%
\begin{flushright}
\textbf{\large{}P2-039}
\par\end{flushright}{\large \par}

\begin{center}
\textbf{\large{}Non-local parallel transport in BOUT++}
\par\end{center}{\large \par}

\begin{center}
\textbf{J.T. Omotani}$^{a}${*}\negthinspace{}, B.D. Dudson$^{b}$\negthinspace{},
E. Havl\'i\v{c}kov\'a$^{a}$ and M. Umansky$^{c}$ 
\par\end{center}

\begin{center}
\textit{$^{a}$CCFE, Culham Science Centre, Abingdon, Oxon OX14 3DB,
UK}
\par\end{center}

\begin{center}
\textit{$^{b}$York Plasma Institute, Department of Physics, University
of York, Heslington, York, YO10 5DD, UK}
\par\end{center}

\begin{center}
\textit{$^{c}$Lawrence Livermore National Laboratory, Livermore,
CA 94550, US}
\par\end{center}
\begin{abstract}
Non-local closures allow kinetic effects on parallel transport to
be included in fluid simulations. This is especially important in
the scrape-off layer, but to be useful there the non-local model requires
consistent kinetic boundary conditions at the sheath. A non-local
closure scheme based on solution of a kinetic equation using a diagonalized
moment expansion has been previously reported. We derive a method
for imposing kinetic boundary conditions in this scheme and discuss
their implementation in BOUT++. To make it feasible to implement the
boundary conditions in the code, we are lead to transform the non-local
model to a different moment basis, better adapted to describe parallel
dynamics. The new basis has the additional benefit of enabling substantial
optimization of the closure calculation, resulting in an $\mathcal{O}(10)$
speedup of the non-local code.
\end{abstract}

\section{Introduction}

Kinetic effects on parallel dynamics may be important in cases where
we would otherwise like to use fluid models: in the scrape-off layer
(SOL), as the collision length is often comparable to the parallel
connection length; or if, for example, we wish to include some Landau
damping physics. To avoid the computational expense of moving to fully
kinetic simulations, we can introduce these kinetic effects into fluid
models through non-local closures that solve (approximately) the electron
kinetic problem in quasi-steady-state. Here we discuss some new developments
to the non-local closure model implemented in BOUT++\citep{dudson2011},
first described in \citep{omotani2013non} and based on the method
of \citep{ji:022312}, in which the 1d kinetic equation is solved
using a moment expansion truncated at very high order (up to several
hundred moments).

In order for a non-local closure to be useful in the SOL it requires
boundary conditions at the sheath edge. These must go beyond just
the fluid velocity (Bohm condition) and heat transmission to specify
completely the boundary conditions for the kinetic equation being
solved. We describe below (Section \ref{sec:Sheath-boundary-conditions})
a method for and implementation of such kinetic boundary conditions
in the simplest case, neglecting secondary electron emission.

The kinetic boundary condition depends only on the parallel velocity,
so the boundary equations are separable. It also introduces a sharp
feature in the distribution function (due the tail absorbed by the
wall being removed), and therefore requires a large number of moments,
but only in the parallel velocity part of the distribution function.
In the previous implementations of this model the moment expansion
used basis functions depending on pitch angle (Legendre polynomials)
and speed (associated Laguerre polynomials). In order to achieve a
certain resolution in the parallel velocity, both of these expansions
must be taken to the same order so that we have $\mathcal{O}(n^{2})$
moments for some $n$. In order to take advantage of the separation
into parallel and perpendicular velocity parts we here reformulate
the closures on a new basis better adapted to the problem at hand,
namely an expansion in parallel velocity (Hermite polynomials) and
perpendicular speed (Laguerre polynomials). In order to resolve the
same features in the parallel velocity we still need $\mathcal{O}(n)$
components in the Hermite expansion, but we can now set the order
of the Laguerre part independently, allowing the total number of moments
to be $\mathcal{O}(n\times m)$ with $m\ll n$. The transformation
to the new basis is presented in Section \ref{sec:Choice-of-moment}.

However, the new basis is not only useful for the boundary conditions.
Since we are solving a 1d kinetic problem, the separation between
parallel and perpendicular velocities is generally a useful one to
make; although the collision operator does couple the parallel and
perpendicular velocity parts, if collisions are the dominant process
we return to the local limit exactly (regardless of the order of the
truncation) and so we need to optimize only for the case when they
are not too strongly coupled. Thus, as we show for the examples in
Section \ref{sec:Comparison}, we can make substantial performance
gains for little loss in accuracy by using the new basis and choosing
the orders of the Hermite and Laguerre expansions appropriately.

\section{Choice of moment basis\label{sec:Choice-of-moment}}

Previous work on this non-local model\citep{ji:022312,omotani2013non}
used a moment basis of Legendre polynomials in pitch angle, $P_{l}(\cos\theta)$,
and associated Laguerre polynomials in speed, $L_{k}^{(l+1/2)}(s^{2})$
(the `old basis'). Throughout we use $\vec{s}=\vec{v}/v_{T}$ (correspondingly
$s=v/v_{T}$, $s_{\|}=v_{\|}/vT$, $s_{\perp}=v_{\perp}/v_{T}$, $s_{\text{sheath}}=v_{\text{sheath}}/v_{T}$)
for velocities normalized by the thermal speed, $v_{T}=\sqrt{\frac{2T}{m}}$.
The old basis is well-adapted for the calculation of the collision
matrix\citep{ji2006exact} which, being isotropic, is block-diagonal
in $l$. However, the calculation of parallel closures is highly anisotropic;
in this case a basis in which parallel velocity, $v_{\|}$, and perpendicular
velocity, $v_{\perp}$, are separable is more natural and convenient
(the `new basis'). This is especially true for the calculation of
sheath boundary conditions (section \ref{sec:Sheath-boundary-conditions}),
which originally motivated the change.

To choose the basis functions explicitly, we identify the appropriate
sets of orthogonal polynomials. For the parallel velocity we take
the Hermite polynomials $H_{p}(s_{\|})$ which are the complete set
of orthogonal polynomials on the interval $(-\infty,\infty)$ with
weight function $e^{-s_{\|}^{2}}$. For the perpendicular velocity
we take the Laguerre polynomials $L_{j}(x)$ which are the complete
set of orthogonal polynomials on the interval $[0,\infty)$ with weight
function $e^{-x}$.

Since $\cos\theta=\frac{s_{\|}}{s}$, $s^{2}=s_{\|}^{2}+s_{\perp}^{2}$
and the $P_{l}$ are odd or even functions according as $l$ is odd
or even, $s^{l}P_{l}(\cos\theta)$ and $L_{k}^{(l+1/2)}(s^{2})$ are
polynomials in $s_{\|}$ and $s_{\perp}^{2}$. They are therefore
given by a finite sum of the new basis functions, 
\begin{align}
s^{l}P_{l}(\cos\theta)L_{k}^{(l+1/2)}(s^{2}) & =\sum_{p=0}^{p_{\text{max}}}\sum_{j=0}^{j_{\text{max}}}T_{lk}^{pj}H_{p}(s_{\|})L_{j}(s_{\perp}^{2})=\sum_{p=0}^{p_{\text{max}}}T_{lk}^{p\hat{j}}H_{p}(s_{\|})L_{\hat{j}}(s_{\perp}^{2})
\end{align}
with $p_{\text{max}}=l+2k$, $\hat{j}=\frac{l-p}{2}+k$. Similarly
\begin{align}
H_{p}(s_{\|})L_{j}(s_{\perp}^{2})= & \sum_{l=0}^{l_{\text{max}}}\sum_{k=0}^{k_{\text{max}}}\left(T^{-1}\right)_{pj}^{lk}s^{l}P_{l}(\cos\theta)L_{k}^{(l+1/2)}(s^{2})=\sum_{l=0}^{l_{\text{max}}}\left(T^{-1}\right)_{pj}^{l\hat{k}}s^{l}P_{l}(\cos\theta)L_{\hat{k}}^{(l+1/2)}(s^{2})
\end{align}
with $l_{\text{max}}=p+2j$, $\hat{k}=\frac{p-l}{2}+j$. To compute
the collision matrix in the new basis exactly up to some order, we
may transform the result in the old basis (${\displaystyle C_{p'j'}^{pj}=\sum_{l,l'=0}^{l_{\text{max}}}\sum_{k,k'=0}^{k_{\text{max}}}T_{lk}^{pj}C_{l'k'}^{lk}\left(T^{-1}\right)_{p'j'}^{l'k'}}$),
with the collision matrix in the old basis only being required at
finite order.

We can now write the kinetic equation in the new moment basis, where
the moments are defined as ${\displaystyle n^{pj}=\frac{1}{2^{p}p!}}\int d^{3}v\, H_{p}(s_{\|})L_{j}(s_{\perp}^{2})f(\vec{v})$
and using the dimensionless length $z$ defined by $\frac{\partial\ell}{\partial z}=\lambda_{C}$,
\begin{align}
v_{\|}\frac{\partial\delta f_{e}}{\partial\ell} & =C\!\left(f_{e}^{(0)}\!+\!\delta f_{e}\right)-v_{\|}\frac{\partial\langle f_{e}^{(0)}\rangle}{\partial\ell}\nonumber \\
\rightarrow\Psi_{p',j'}^{p,j}\frac{\partial n^{p',j'}}{\partial z} & =C_{p',j'}^{p,j}n^{p',j'}+g^{p,j}
\end{align}
The moments of the free-streaming operator are straightforward to
compute in the new basis
\begin{align}
\Psi_{p',j'}^{p,j} & \equiv\int d^{3}v\, H_{p}(s_{\|})L_{j}(s_{\perp}^{2})\sum_{p',j'}\frac{1}{\pi^{3/2}v_{T}^{3}}e^{-s^{2}}s_{\|}H_{p'}(s_{\|})L_{j'}(s_{\perp}^{2})\nonumber \\
 & =2^{p}p!\left(\frac{1}{2}\delta_{p,p'+1}+p'\delta_{p,p'-1}\right)\delta_{j,j'}
\end{align}
Having found the coefficients of the moment equations we can, as before\citep{omotani2013non},
diagonalize the system by going to an eigenvector basis and compute
the closures as sums of integrals.

There is one slight complication. To compute the closures, we must
remove from the system the equations for density, fluid velocity and
temperature (which are solved dynamically). However, the moment corresponding
to the temperature in the new basis is a linear combination of the
$(p=2,j=0)$ and $(p=0,j=1)$ moments, so we must apply a further
transformation to this pair of moments to remove the temperature part
from the closures. We call this transformation $R$, and solve for
the set of moments $n_{*}^{pj}=R_{p'j'}^{pj}n^{p'j'}$. $R$ differs
from the identity only in four components, which are
\begin{align}
\left(\begin{array}{cc}
R_{0,1}^{0,1} & R_{2,0}^{0,1}\\
R_{0,1}^{2,0} & R_{2,0}^{2,0}
\end{array}\right) & =\left(\begin{array}{cc}
\frac{1}{2} & -1\\
\frac{1}{3} & \frac{4}{3}
\end{array}\right)
\end{align}

\section{Comparison of bases\label{sec:Comparison}}

\begin{wrapfigure}{O}{0.5\columnwidth}%
\includegraphics[width=7.5cm]{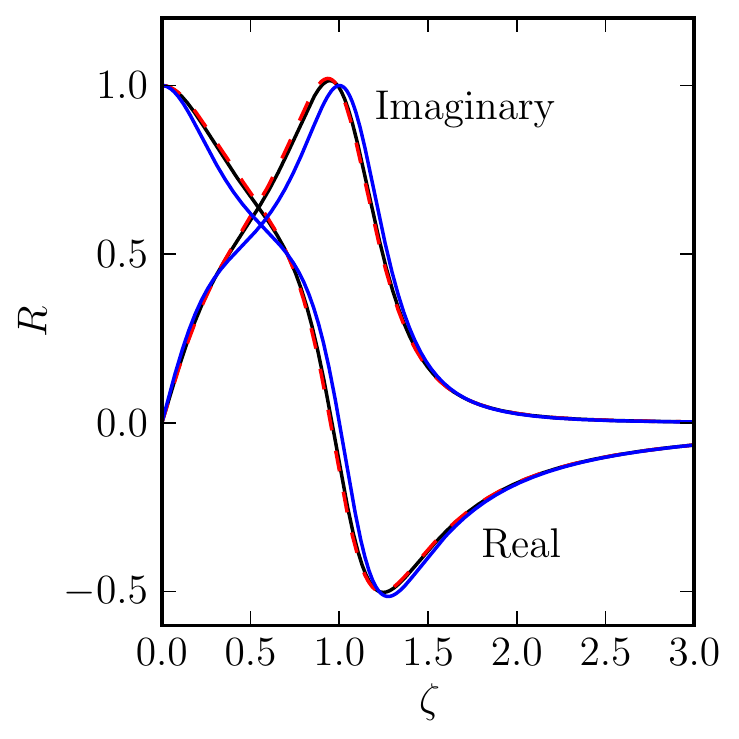}\caption{\label{fig:Landau-damping}Response function for Hammett-Perkins closure
(black), old basis with $30\times30$ moments (red, dashed) and new
basis with $30\times2$ moments (blue)}
\end{wrapfigure}%
Landau damping is an interesting test case for these closures because
we have a known collisionless limit from the results of Hammett \&
Perkins\citep{hammett_1990}. It is possible to reproduce this collisionless
limit by replacing the Hammett-Perkins expression for the heat-flux
with the result from the non-local closures being discussed here and
taking a sufficiently small (but non-zero) collisionality. To match
the collisionless limit using the old basis requires $30\times30$
moments for convergence. In the new basis, however, we have the advantage
of being able to much reduce the number of moments; using only $30\times2$
moments gives only a small loss in accuracy, as shown in Figure \ref{fig:Landau-damping}.
Here and below we describe the number of moments in particular cases
as pairs of numbers: (order of Legendre expansion)$\times$(order
of associated Laguerre expansion) for the old basis and (order of
Hermite expansion)$\times$(order of Laguerre expansion) for the new
basis.

\begin{figure}[h]
\begin{minipage}[t]{0.49\columnwidth}%
\includegraphics[width=7.5cm]{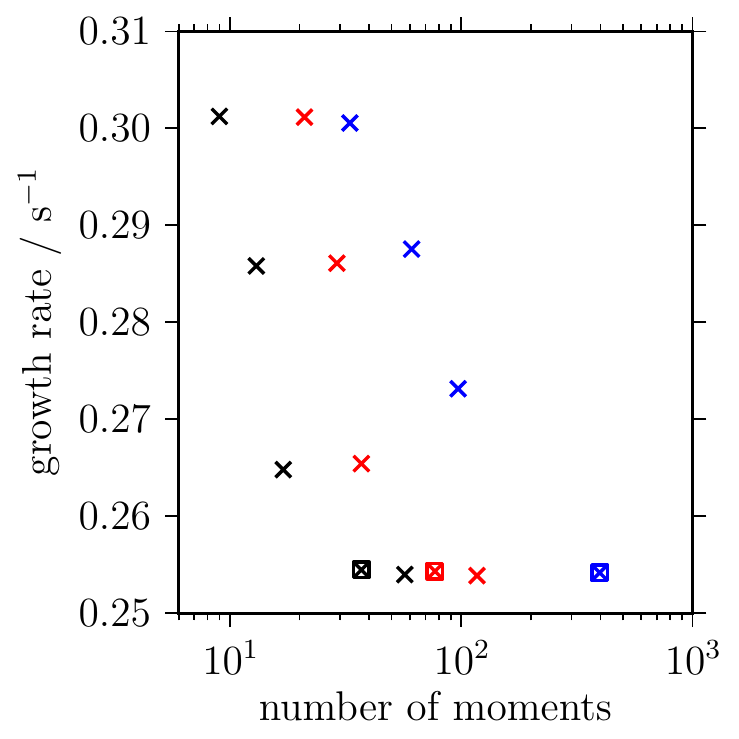}\caption{\label{fig:drift-wave-convergence}Convergence of drift instability
simulation, growth rates for various numbers of moments: new basis
$n\times2$ (black), $n\times4$ (red) and old basis (blue) }
\end{minipage}\hfill{}%
\begin{minipage}[t]{0.49\columnwidth}%
\includegraphics[width=7.5cm]{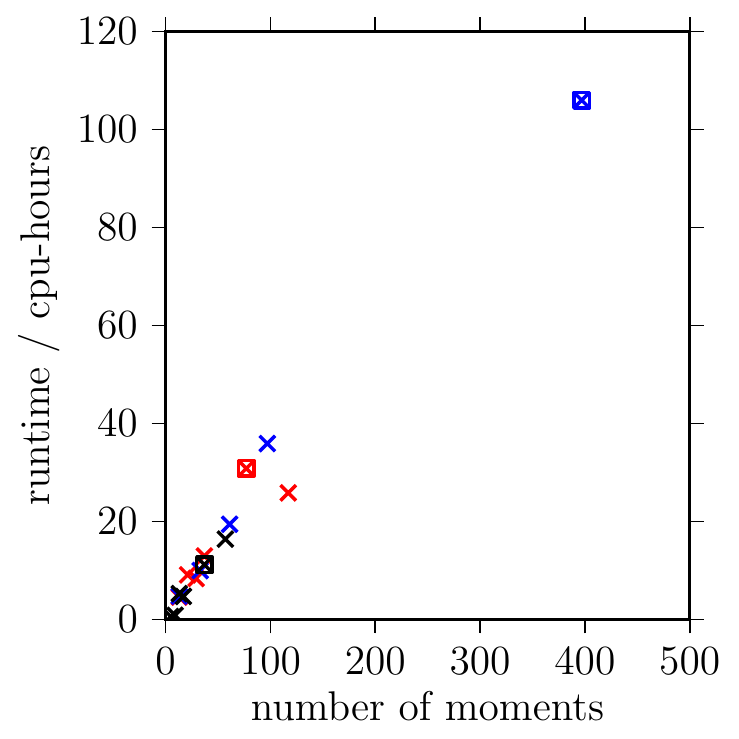}

\caption{\label{fig:drift-wave-run-times}Run times for drift-wave simulation:
new basis 2 (black), $n\times4$ (red) and old basis (blue). Squares
mark the smallest well converged simulations (see Figure \ref{fig:drift-wave-convergence})}
\end{minipage}
\end{figure}

The decrease in the number of moments needed for convergence represents
a significant gain for the performance of the code. To illustrate
this we consider the drift-wave instability test-case, previously
discussed in \citep{omotani_eps2013} but here with the electron parallel
viscosity included. Convergence in the old basis requires $20\times20$
moments, while in the new basis it requires only $20\times2$; the
total run time (for otherwise identical simulations) reduces from
106 cpu-hours to 11 cpu-hours. The simulations were run on a 4-core
desktop machine, using a $32\times32$ grid. The perturbation is seeded
with a wavenumber $k_{\|}=8.15\times\frac{2\pi}{\lambda_{C}}$ so
that we consider a low collisionality case. Figure \ref{fig:drift-wave-convergence}
shows that convergence can be achieved for a much smaller number of
moments using the new basis. The performance gain is demonstrated
by Figure \ref{fig:drift-wave-run-times} where we see that the total
run time for the simulations is directly proportional to the number
of moments used, as the calculation of the closures dominates the
computation time here (although in a typical three-dimensional simulation
there would be other computationally intensive operations, such as
Laplacian inversion, that might be comparable in computational time).

\section{Sheath boundary conditions\label{sec:Sheath-boundary-conditions}}

Calculation of correct sheath boundary conditions is much more complicated
for the non-local model than for simple fluid models, not least because
boundary conditions are required for several hundred moments rather
than just a few. To derive boundary conditions for the non-local model
we start from the simplest possible kinetic sheath boundary condition
(with no secondary electron emission). Considering the sheath where
outgoing $v_{\|}$ is positive
\begin{align}
f_{\text{sheath}}(v_{\|},v_{\perp}) & =\begin{cases}
f_{+}(v_{\|},v_{\perp}) & v_{\|}>0\\
f_{+}(-v_{\|},v_{\perp}) & -v_{\text{sheath}}<v_{\|}<0\\
0 & v_{\|}<-v_{\text{sheath}}
\end{cases}
\end{align}
where $v_{\text{sheath}}=\sqrt{-\frac{2e\phi_{\text{sheath}}}{m_{e}}}$
is the speed needed to cross the sheath potential. As this boundary
condition is independent of $v_{\perp}$, the calculation is much
cleaner in the new basis, since the expansion in $L_{j}(s_{\perp}^{2})$
is trivial everywhere.

First we translate this boundary condition into the moment representation.
\begin{align*}
n^{p,j} & \equiv\frac{1}{2^{p}p!}\int d^{3}v\, H_{p}(s_{\|})L_{j}(v_{\perp}^{2})f(v_{\|},v_{\perp})
\end{align*}
\begin{align}
2^{p}p!\left.n^{p,j}\right|_{\text{sheath}} & =\int d^{2}v_{\perp}\int_{0}^{\infty}dv_{\|}\, H_{p}(s_{\|})L_{j}(s_{\perp}^{2})f_{+}(v_{\|},v_{\perp})\nonumber \\
 & \quad+\int d^{2}v_{\perp}\int_{-s_{\text{sheath}}}^{0}dv_{\|}\, H_{p}(s_{\|})L_{j}(s_{\perp}^{2})f_{+}(-v_{\|},v_{\perp})\nonumber \\
 & =\begin{cases}
\begin{array}{l}
2\int d^{2}v_{\perp}\int_{0}^{\infty}dv_{\|}\, H_{p}(s_{\|})L_{j}(s_{\perp}^{2})f_{+}(v_{\|},v_{\perp})\\
-\int d^{2}v_{\perp}\int_{s_{\text{sheath}}}^{\infty}dv_{\|}\, H_{p}(s_{\|})L_{j}(s_{\perp}^{2})f_{+}(v_{\|},v_{\perp})
\end{array} & p\text{ even}\\
\int d^{2}v_{\perp}\int_{s_{\text{sheath}}}^{\infty}dv_{\|}\, H_{p}(s_{\|})L_{j}(s_{\perp}^{2})f_{+}(v_{\|},v_{\perp}) & p\text{ odd}
\end{cases}\label{eq:integral-boundary-condition}
\end{align}
as $H_{p}(-v_{\|})=(-1)^{p}H_{p}(v_{\|})$. Since $f_{+}(v_{\|},v_{\perp})\equiv f(v_{\|},v_{\perp})$
for $v_{\|}>0$ we can expand $f_{+}$ in moments
\begin{align}
\left.n^{p,j}\right|_{\text{sheath}} & =\begin{cases}
n^{p,j}+\sum_{p'}\left(2X_{p'}^{p}n^{(2p'+1),j}-\beta_{p}^{p'}n^{(2p'+1),j}-\gamma_{p'}^{p}n^{2p',j}\right) & p\text{ even}\\
\sum_{p'}\left(\alpha_{p'}^{p}n^{(2p'+1),j}+\beta_{p'}^{p}n^{2p',j}\right) & p\text{ odd}
\end{cases}\label{eq:matrix-boundary-conditions}
\end{align}
defining
\begin{align}
X_{p'}^{p} & =\frac{1}{2^{p}p!}\int_{0}^{\infty}ds_{\|}\, H_{p}(s_{\|})H_{(2p'+1)}(s_{\|})e^{-s_{\|}^{2}}\nonumber \\
\alpha_{p'}^{p} & =\frac{1}{2^{p}p!}\int_{s_{\text{sheath}}}^{\infty}ds_{\|}\, H_{(2p+1)}(s_{\|})H_{(2p'+1)}(s_{\|})e^{-s_{\|}^{2}}\nonumber \\
\beta_{p'}^{p} & =\frac{1}{2^{p}p!}\int_{s_{\text{sheath}}}^{\infty}ds_{\|}\, H_{(2p+1)}(s_{\|})H_{2p'}(s_{\|})e^{-s_{\|}^{2}}\nonumber \\
\gamma_{p'}^{p} & =\frac{1}{2^{p}p!}\int_{s_{\text{sheath}}}^{\infty}ds_{\|}\, H_{2p}(s_{\|})H_{2p'}(s_{\|})e^{-s_{\|}^{2}}
\end{align}
 We could use either relation in (\ref{eq:matrix-boundary-conditions})
to determine the odd-$p$ moments in terms of the even-$p$ moments
or vice versa. They must be equivalent (before truncation) but the
odd-$p$ version is simpler, so we use that.

Finally, the boundary condition that we want is on the eigenvector-basis
moments. We need to determine the positive-eigenvalue (outgoing) moments
(and the fluid velocity) in terms of the negative-eigenvalue (incoming)
moments.
\begin{align}
nS_{\|}\delta_{0}^{p}\delta_{0}^{j}+\sum_{B}\mat{W_{-}}{(2p+1),j}B(\hat{n}_{-}^{B}-\hat{n}_{+}^{B}) & =\sum_{qB}\alpha_{q}^{p}\left(\mat{W_{-}}{(2q+1),j}B\left(\hat{n}_{-}^{B}-\hat{n}_{+}^{B}\right)+nS_{\|}\delta_{0}^{q}\delta_{0}^{j}\right)\nonumber \\
 & \quad+\sum_{qrj'B}\beta_{q}^{p}\left(R^{-1}\right)_{2r,j'}^{2q,j}\left(\mat{W_{-}}{2r,j'}B\left(\hat{n}_{-}^{B}+\hat{n}_{+}^{B}\right)+n\delta_{0}^{r}\delta_{0}^{j'}\right)\nonumber \\
\Rightarrow\hat{n}_{+}^{\tilde{B}} & =\sum_{pj}\left(A^{-1}\right)_{p,j}^{\tilde{B}}\left(\sum_{C}B_{C}^{p,j}\hat{n}_{-}^{C}+\sum_{q}\beta_{q}^{p}\left(R^{-1}\right)_{0,0}^{2q,j}n\right)\\
 & \equiv\sum_{C}E_{C}^{\tilde{B}}\hat{n}_{-}^{C}+E_{(0)}^{\tilde{B}}n
\end{align}
where $\mat{W_{\pm}}{p,j}B$ are the matrices of eigenvectors with
positive/negative eigenvalues, $n_{+}^{\tilde{B}}\equiv\left\{ nS_{\|},\hat{n}_{+}^{B}\right\} ^{\tilde{B}}$,
$\mat{W_{(b)}}{(2p+1),j}{\tilde{B}}\equiv\left\{ -\delta_{0}^{p}\delta_{0}^{j},\mat{W_{-}}{(2p+1),j}B\right\} _{\tilde{B}}$
and	
\begin{align*}
A_{\tilde{B}}^{p,j} & =-\mat{W_{(b)}}{(2p+1),j}{\tilde{B}}+\sum_{q}\left(\alpha_{q}^{p}\mat{W_{(b)}}{(2q+1),j}{\tilde{B}}+\sum_{rj'}\beta_{q}^{p}\left(R^{-1}\right)_{2r,j'}^{2q,j}\mat{W_{-}}{2r,j'}{\tilde{B}}\right)\\
B_{C}^{p,j} & =-\mat{W_{-}}{(2p+1),j}C+\sum_{q}\left(\alpha_{q}^{p}\mat{W_{-}}{(2q+1),j}C+\sum_{r,j'}\beta_{q}^{p}\left(R^{-1}\right)_{2r,j'}^{2q,j}\mat{W_{-}}{2r,j'}C\right)
\end{align*}

This boundary condition depends on the value of the sheath potential,
which must be determined self-consistently by imposing a boundary
condition on the current. In the BOUT++ code two options have been
implemented: zero current at either sheath (floating walls) and zero
net current (equal potential at both walls). We compute the sheath
potentials that satisfy the condition on the current by Newton iteration.
To find $E_{C}^{\tilde{B}}$ and $E_{(0)}^{\tilde{B}}$ at each step
of the iteration we interpolate from a set of stored values, pre-computed
for a suitable range of $s_{\text{sheath}}$ (here from 0 to 3 in
steps of 0.01).

\begin{wrapfigure}{o}{0.5\columnwidth}%
\includegraphics[width=7.5cm]{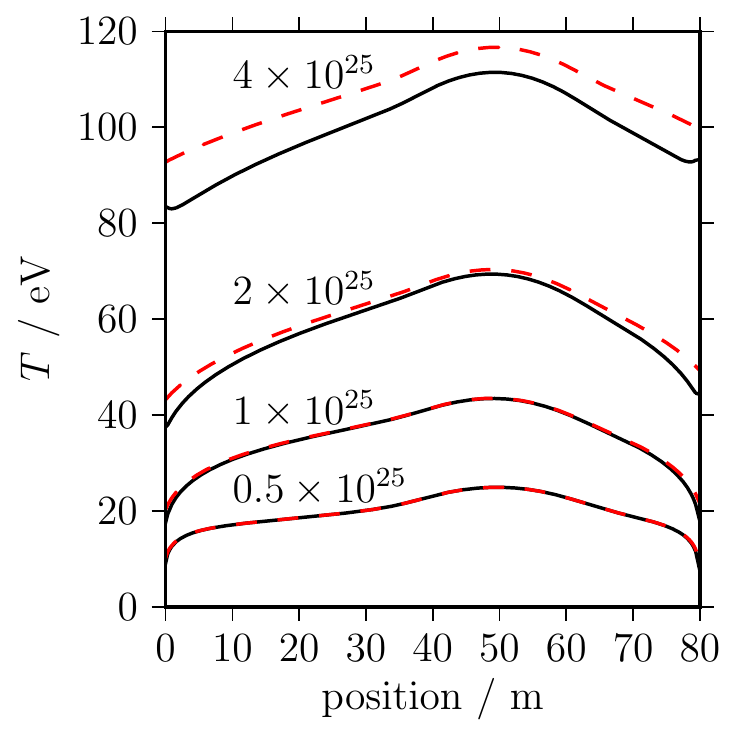}\caption{\label{fig:steady-state-sol-profiles}Steady state profiles for 1d
SOL simulations using sheath boundary (black) conditions original
boundary conditions (red) for several different heat-source amplitudes
(in eV/m$^{3}$/s). Number of moments used is $40\times4$}
\end{wrapfigure}%
In the limit of short collision length, the non-local model asymptotes
to local, collisional (Braginskii) fluid closures. In this case the
influence of the boundary conditions does not propagate into the domain
and it suffices to have the correct fluid velocity and heat-flux at
the sheath. Thus when the electron temperature is low enough the sheath
boundary conditions described here give the same results as the old
ones (used in \citep{omotani2013non}, which impose the correct heat-flux
but do not otherwise enforce sheath boundary conditions on the non-local
model). However, as the temperature (and hence the collision length)
increases it becomes important to use fully correct boundary conditions,
as we see in Figure (\ref{fig:steady-state-sol-profiles}) where the
temperature profiles in steady-state in a one-dimensional SOL model
(see \citep{omotani2013non} for details) are shown. The only parameter
changed is the amplitude of the heat source for the electrons which
is used to vary the electron temperature. For low temperatures (up
to $\sim$50eV here, corresponding to a collision length of $\sim$4m)
both methods give the same results, but at higher temperatures ($\sim$100eV
corresponding to a collision length of $\sim$14m) we can see that
the details of the boundary conditions have a significant effect on
the results.

\section{Conclusions}

A scheme to give kinetic sheath boundary conditions for non-local
parallel closures has been derived and implemented in BOUT++, allowing
kinetic effects to be consistently included in fluid models of the
SOL. This opens up a much wider parameter space to investigate the
behaviour of the SOL plasma through three-dimensional fluid simulations,
as these can now be extended to low collisionality.

The change in moment basis also gives an $\mathcal{O}(10)$ speed-up
in the evaluation of the closures for typical parameters, making three-dimensional
simulations using the non-local code much more readily practicable.

Future work will investigate the extension of the boundary conditions
to include the effects of secondary electron emission and begin to
apply these non-local closures to three-dimensional SOL simulations,
initially focusing on filament dynamics.

\section*{Acknowledgements}

This work was funded by the RCUK Energy Programme {[}under grant EP/I501045{]}.
To obtain further information on the data and models underlying this
paper please contact PublicationsManager@ccfe.ac.uk. The views and
opinions expressed herein do not necessarily reflect those of the
European Commission.

\bibliographystyle{unsrt}
\bibliography{references}

\end{document}